\documentclass[aps,prl,twocolumn,showpacs,superscriptaddress,groupedaddress]{revtex4}

\usepackage{bbm}

\usepackage{graphicx}
\usepackage{dcolumn}
\usepackage{amsmath}
\usepackage{bm}
\usepackage{color}
\usepackage{mathrsfs}
\usepackage{epstopdf}
\usepackage{amssymb}
\usepackage{subfigure}
\usepackage{enumitem}
\usepackage{multirow}
\usepackage{exscale}
\usepackage{relsize}
\usepackage{dsfont}

\usepackage{float}

\newcommand{\be}{\begin{equation}}
\newcommand{\ee}{\end{equation}}
\newcommand{\bey}{\begin{eqnarray}}
\newcommand{\eey}{\end{eqnarray}}
\newcommand{\bw}{\begin{widetext}}
\newcommand{\ew}{\end{widetext}}

\newcommand{\ba}{\begin{array}}
\newcommand{\ea}{\end{array}}
\newcommand{\bi}{\begin{itemize}}
\newcommand{\ei}{\end{itemize}}
\newcommand{\bem}{\begin{enumerate}}
\newcommand{\eem}{\end{enumerate}}

\begin{document}

\title{Universal scaling of work statistics in conformal-field-theory models}

\author{Zhaoyu Fei} \email[Email: ]{1501110183@pku.edu.cn}
\affiliation{Graduate School of China Academy of Engineering Physics,
No. 10 Xibeiwang East Road, Haidian District, Beijing, 100193, China}

\author{C. P. Sun} \email[Email: ]{suncp@gscaep.ac.cn}
\affiliation{Graduate School of China Academy of Engineering Physics,
No. 10 Xibeiwang East Road, Haidian District, Beijing, 100193, China}
\affiliation{Beijing Computational Science Research Center, Beijing, 100193, China}

 \date{\today}

\begin{abstract}
In this paper, we systematically study the work statistics for quantum phase transition. For a quantum system approached by an anisotropic conformal field theory near the critical point, the driving protocols is divided into three different regimes for different quench rates,  which reflects the competition between the frozen time and the quench time scale. In each regime, we find universal scaling behaviors in work statistics (after renormalization). It is shown that the critical exponents are determined by the space-time dimension $d$, the dynamical critical exponent $z$, the correlation-length exponent $\nu$, and the power-law protocols. These universal scalings in nonequilibrium process may be found in quantum phase transition by measuring the Loschmidt echo or  the Ramsey  interferometry.
\end{abstract}

\maketitle

\emph{Introduction}---In the past decades, quantum quench across the critical point of quantum phase transition attracted much attention both in theories and experiments~\cite{subir2011, greiner2002, baumann2011, wzurek2005, dy2010, no2011}, since universal scaling behavior may arise during  nonequilibrium  processes. Surprisingly, the recent studies have provided a panoramic view of the process: from extremely slow to extremely fast quench rate~\cite{qu2016,old2016,an2017}. Basically, the protocols is divided into three different regimes according to different quench rates. In the slow quench regime, the creation of excitations (topological defects) 
is usually described by the Kibble-Zurek mechanism~\cite{kibble1976,zurek1985}
which uncovers the nonadiabatic effect near the critical point. 
In the fast regime, recent
holographic studies~\cite{buchel2012} also revealed interesting new scaling behavior of the renormalized quantities, which is later shown to be universal for quantum field theories flowing from an ultraviolet (UV) fixed point (described by the conformal field theory)~\cite{un2014,un2015,sm2015}. In the instantaneous regime, it was argued~\cite{cala2006} that the universal relaxation process to a critical Hamiltonian from a noncritical one. Importantly, it was explicitly shown in~\cite{qu2016,an2017} that scaling behavior of (renormalized) quantities smoothly interpolates between different regimes with one quench protocol in free scalar and fermion field theories.

The above studies mainly focused on the scaling behavior of the expectation of observables. Nevertheless, due to quantum uncertainty, quantum fluctuations characterized by the cumulants of the excitations~\cite{delcampo2018} and the trajectory work~\cite{wo2020} also exhibit universal scaling behaviors. In the two-point measurement scheme, the trajectory work is defined as the energy difference between the initial and final projective measurements~\cite{aq2000,ja2000,flu2007,esposito2009}. As a result, the characteristic function of work [the Fourier transform of the work
distribution $P(w)$] $\chi(u)$ reads
\be
\chi(u)=\int\mathrm dw P(w)e^{iuw}=\mathrm{Tr}[e^{iu\hat H^{\mathrm{H}}(t_f)}e^{-iu\hat H(t_i)}\hat\rho],
\ee
where $\hat \rho$ denotes the initial state, $\hat H(t_i)$ and $\hat H^{\mathrm{H}}(t_f)$ denote Hamiltonians in the Heisenberg picture corresponding to the initial time $t_i$ and final time $t_f$ of the quench. Then, the $n$th cumulant of work $\kappa_n$ is defined as the $n$th-order derivative of $\ln\chi(u)$, $\kappa_n=\partial^n\ln\chi(u)/\partial (iu)^n$. In analog to the partition function encoding essential information about an equilibrium state, the work statistics encodes essential information about the fluctuations in the nonequilibrium process. It allows us to understand the emergence of irreversibility in stochastic thermodynamics (via the fluctuation
relations~\cite{sekimoto2010,jarzynski2011,seifert2012,esposito2009,dorner2012}). Meanwhile, it is related to other interesting quantities employed to study the nonequilibrium process for quantum many-body systems like the Loschmidt echo~\cite{st2008,de2006},  the Ramsey  interferometry~\cite{dorner2013} and the dynamical quantum phase transition~\cite{dy2013,qua2016,wo2020}.

In this Letter, we present a panoramic description of the universal scaling behavior of the work cumulants  in all the three regimes, which implies the competition between the frozen time (Kibble-Zurek mechanism) and the quench time scale. The symmetries of the anisotropic conformal field theory near the critical point make sure the universality of the scaling behaviors. Compared to previous results~\cite{wo2020} based on the quasiparticle picture, our results are obtained by using the dimensional analysis and thus universal. Moreover, we give a clear description of the renormalization procedure in nonequilibrium thermodynamics. Actually, similar to the \emph{minimal subtraction scheme}, the divergent part of quantities is renormalized by subtracting its value in the adiabatic limit (sudden quench limit) for the UV renormalization (infrared (IR) renormalization). And, the higher order corrections (cut-off dependent) may arise in some cases. Finally, our predictions are explicitly verified in an exactly solvable model: a scalar field with changing mass.

\emph{Quantum quench in quantum phase transition}---In (second-order) quantum phase transition, the energy gap $E_g$, the relaxation time $\tau$, and the correlation length $\xi$~\cite{subir2011, dy2010} scale as
\be
\label{e1}
E_g\sim|\lambda|^{z\nu},\tau\sim|\lambda|^{-z\nu},\xi\sim|\lambda|^{-\nu},
\ee
where $\lambda$ measures the distance from the critical point. If $\lambda$ is controlled according to a time-dependent protocol which starts from $t=t_i$ and ends at $t=t_f$ and changes over a time scale $\delta$ (see Eqs.~(\ref{e2}, \ref{e3})), we call this quantum  quench when the initial state is the vacuum, regardless of the rate of change~\cite{old2016}. We only consider global quench which keeps the space-translation symmetry in the following. Also, we assume the quench process passes through or approaches a critical point at $t=0$ and the protocol exhibits the power-law behavior near the critical point,
\be
\label{e2}
\lambda(t)\sim\left(t/\delta\right)^r.
\ee
Without loss of generality, it follows from Eq.~(\ref{e1}, \ref{e2}) that the  time-dependent energy-gap protocol $E_g(t)$ reads
\be
\label{e3}
E_g(t)=E_0f(t/\delta),
\ee
where $E_0$ denotes the energy scale,  $f(x)\to |x|^{z\nu r}$ when $|x|\to 0$. We also assume $f(x)$ is up to $O(1)$ all the time.

\begin{figure}[t]
\centering
\includegraphics[width=0.48\textwidth]{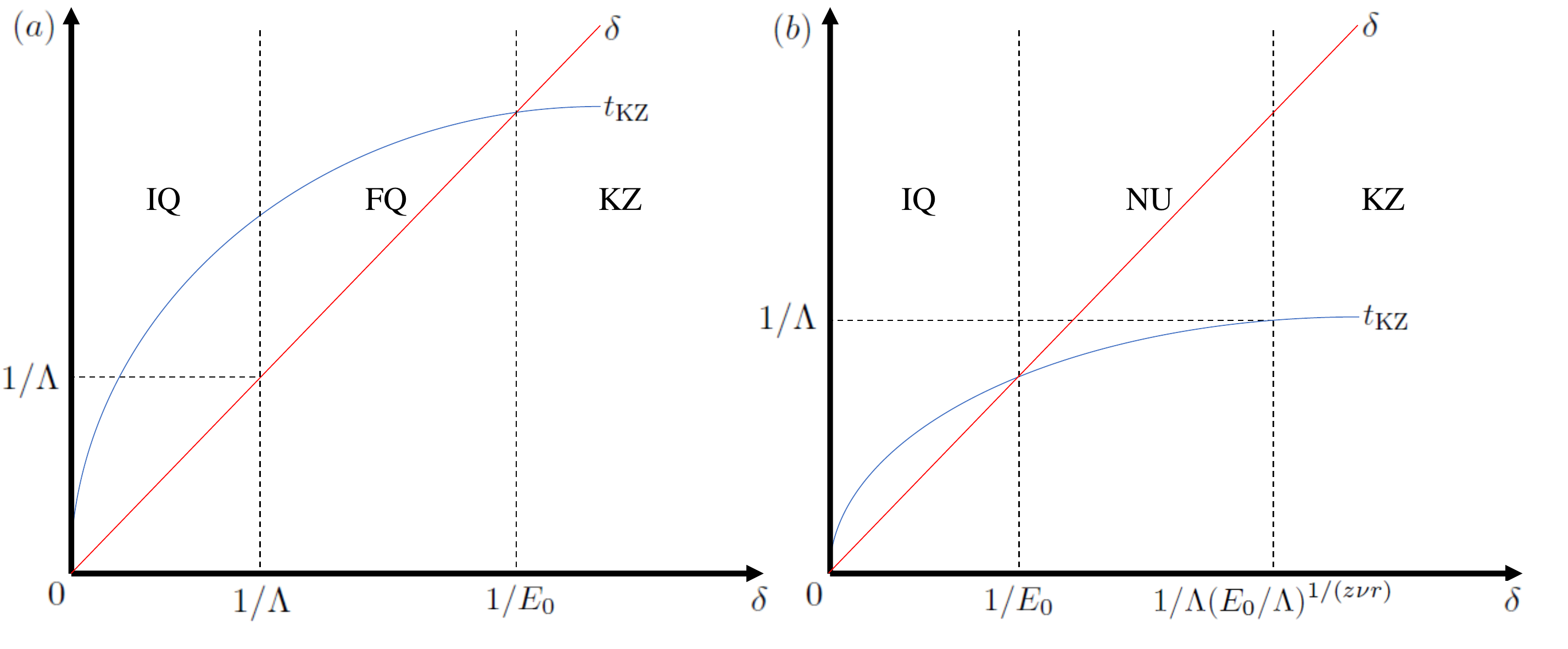}
\caption{Two cases for quenches in quantum phase transition. (a) $E_0\ll\Lambda$. Instantaneous quench (IQ) regime, fast quench (FQ) regime, and Kibble-Zurek (KZ) regime. $t_{\text{KZ}}\propto E_0^{-1}(E_0\delta)^{1/(z\nu r)}$. (b) $E_0>\Lambda$. Instantaneous quench (IQ) regime, non-universal (NU) regime, and Kibble-Zurek (KZ) regime.}\label{fig1}
\end{figure}

Let $\Lambda$ denote the ultraviolet (UV) cut-off scale. Then, according to different quench rates, the protocols can be divided into three different regimes in two cases: $E_0\ll\Lambda$ or $E_0>\Lambda$ (Fig.~\ref{fig1}). We only discuss the case $E_0\ll\Lambda$ in the following paper (for the case $E_0>\Lambda$, see~\cite{foot1}).

\begin{figure}[t]
\centering
\includegraphics[width=0.40\textwidth]{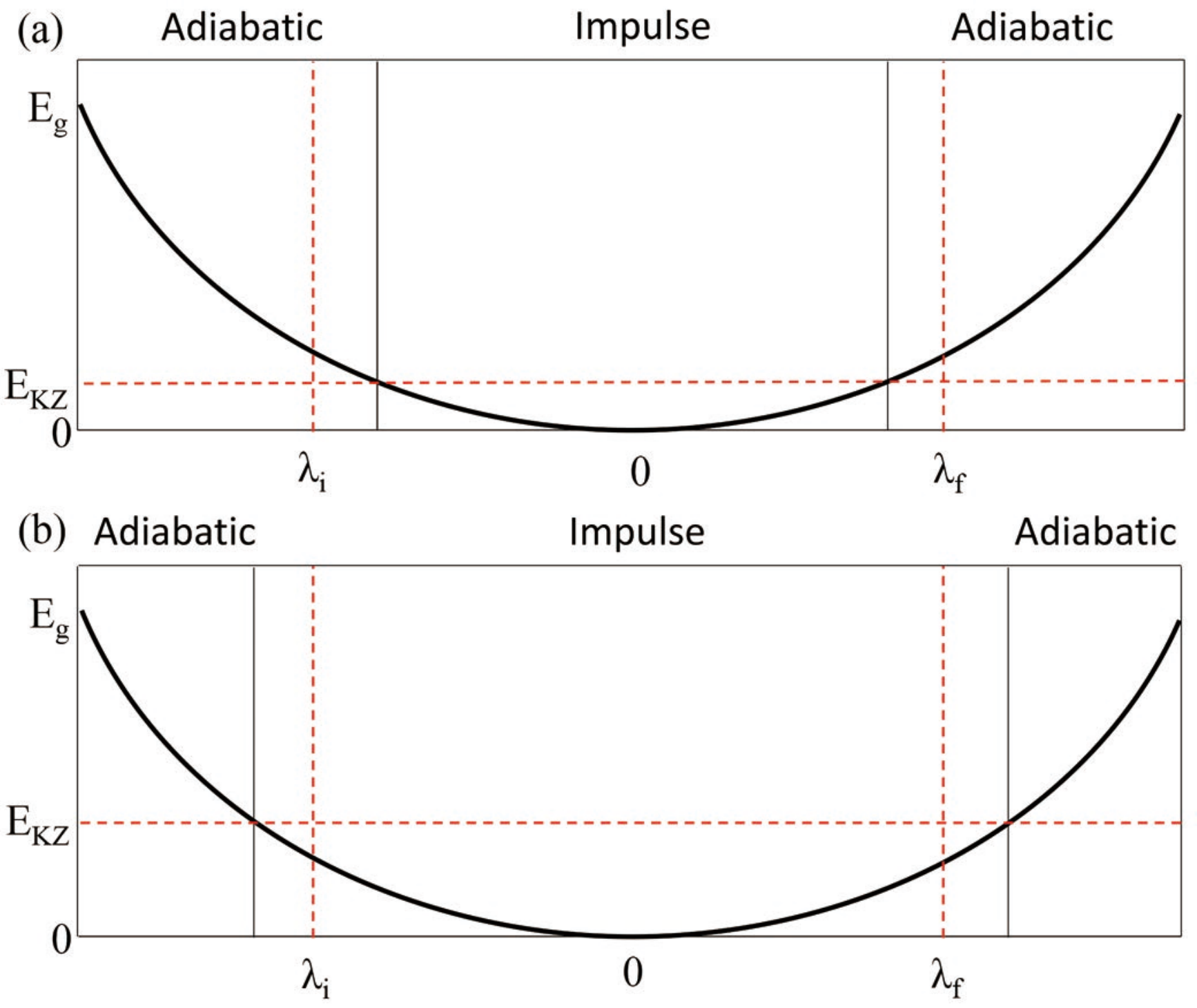}
\caption{The competition between the frozen time $t_{KZ}$ and the quench time scale $\delta$ in the adiabatic-impulse-adiabatic stages. The black solid line denotes the energy gap $E_g$. (a) the Kibble-Zurek regime. A part of the protocol is inside the impulse stage ($t_{KZ}<\delta$). (b) the fast quench regime. All the protocol is inside the impulse stage ($t_{KZ}>\delta$). }\label{fig3}
\end{figure}

\emph{A: Kibble-Zurek regime}---When the quench rate is slow compared to the initial gap ($ E_0 \delta\gg 1$), many systems show Kibble-Zurek (KZ) scaling~\cite{kibble1976,zurek1985,dy2010,un2005,no2011,le2010}. The quench process can be approximated by the adiabatic-impulse-adiabatic stages (see Fig.~(\ref{fig3}a)). Because the quench rate is slow, the quench process is adiabatic unless the adiabatic condition
\be
\label{e4}
\frac{1}{E_g(t)^2}\frac{\mathrm dE_g(t)}{\mathrm dt}\ll 1
\ee
is broken near the critical point (i.e., in the impulse stage) due to the zero energy gap. The boundaries of the impulse stage are at times $t = \pm t_{\text{KZ}}$ (frozen time) which follows from Eqs.~(\ref{e3}, \ref{e4}) reads
\be
\label{e5}
t_{\text{KZ}}\propto E_0^{-1}\left(E_0\delta\right)^{\frac{z\nu r}{1+z\nu r}}.
\ee
Then, the corresponding correlation length $\xi_{\text{KZ}}$ and  energy gap $E_{\text{KZ}}$ read
\be
\label{e6}
\xi_{\text{KZ}}\sim|\lambda(t_{\text{KZ}})|^{-\nu}\sim \delta^{\frac{\nu r}{1+z\nu r}},
\ee
\be
\label{e7}
E_{\text{KZ}}=E_g(t_{\text{KZ}})\sim \delta^{-\frac{z\nu r}{1+z\nu r}}.
\ee
If the unique length scale in the impulse stage is $\xi_{\text{KZ}}$, expectation value of given operator $\hat O_{\Delta}$ (which are independent of $\lambda$) scales as powers of $\xi_{\text{KZ}}$~\cite{old2016}
\be
\label{e8}
\langle O_\Delta\rangle_{\text{re}}\sim \xi_{\text{KZ}}^{-\Delta}
\ee
where the subscript ``re'' denote quantities after renormalization (see the next section),  and $\Delta$ is the scaling dimension of the operator $\hat O_{\Delta}(t)$. In another word
\be
\label{e9}
\langle O_\Delta\rangle_{\text{re}}\sim \delta^{-\frac{\Delta\nu r}{1+z\nu r}}.
\ee

The energy of the final state $\langle H(t)\rangle_r$ should be proportional to the final energy gap. When the process ends in the adiabatic stage, $E_g(t)\sim E_0$, we have
\be
\label{e10}
\langle H(t)\rangle_{\text{re}}\propto V_{d-1}E_0\xi_{\text{KZ}}^{-(d-1)}\sim \delta^{-\frac{(d-1)\nu r}{1+z\nu r}}
\ee
by using the dimensional analysis, where $d$ is the space-time dimension, $V_{d-1}$ is the volume of the system. Similarly for work statistics~\cite{wo2020}, the $n$th-order cumulant work moment $\kappa_n$ is obtained as
\be
\label{e11}
(\kappa_n)_{\text{re}}\propto V_{d-1}E_0^n\xi_{\text{KZ}}^{-(d-1)}\sim \delta^{-\frac{(d-1)\nu r}{1+z\nu r}}.
\ee
When the process ends in the impulse stage, $E_g(t)\sim E_{\text{KZ}}$, we have
\be
\label{e12}
(\kappa_n)_{\text{re}}\propto V_{d-1}E_{\text{KZ}}^n\xi_{\text{KZ}}^{-(d-1)}\sim \delta^{-\frac{(d-1+nz)\nu r}{1+z\nu r}}.
\ee
It is emphasized that Eqs.~(\ref{e11}, \ref{e12}) are obtained by using the dimensional analysis regardless of the quasi-particle picture in Ref.~\cite{wo2020}. Moreover, the KZ scaling is closely related to the conformal field theory where the exponents $z, \nu, \delta$ imply the scale symmetry of the field.

\emph{B: Fast quench}---As the quench time scale $\delta$ increases,  both $E_{KZ}$ and the range of the impulse stage increase. When the all the quench process is included in the impulse stage  ($t_{KZ}>\delta$), the time scale for the nonadiabatic effect changes from $t_{KZ}$ to $\delta$  and the KZ scaling is invalid (see Fig.~\ref{fig3}b). Here, this regime is called the fast quench (FQ) regime, $E_0\delta\ll1\ll\Lambda\delta$, and we show new scaling behavior in the following.

Because the systems is always near the critical point in the fast quench regime, we evaluate quantities by using perturbation expansion. In particular, consider a generic action near the critical point
\be
\label{e13}
S[\hat \phi,\hat O_\Delta] = S_{\text{CFT}}[\hat \phi,\hat O_\Delta]-\int\mathrm dt\int\mathrm d^{d-1}x \lambda(t) \hat O_\Delta(\bm x,t).
\ee
Here $\lambda(t)=\lambda_0h(t/\delta)$ is the protocol which starts from $\lambda_{i}=\lambda_0h(t_i/\delta)$ and ends at $\lambda_{f}=\lambda_0h(t_f/\delta)$, $\hat O_\Delta(\bm x,t)$ is in the Heisenberg picture, $S_{\text{CFT}}$ is the anisotropic conformal field theory  action (which has scale, translation and spatial rotation symmetries)~\cite{no2008} describing the UV fixed point, and $\hat O_\Delta$ is a
relevant operator ($\Delta<d$) with the scaling dimension $\Delta$. We assume $g(x)$ is up to $O(1)$ all the time for simplicity.
Then, according to Refs.~\cite{un2014,un2015}, $\langle O_\Delta(t)\rangle$ is calculated by using linear response theory ($\hbar=1$) as
\begin{gather}
 \begin{split}
 \label{e14}
\langle O_\Delta(\bm x,t)\rangle=&\langle O_\Delta(\bm x)\rangle_{\lambda_i}-i\lambda_0\int_{t_i}^{t}\mathrm dt'\int\mathrm d^{d-1}x' h(t'/\delta)\\
&\times\langle [\hat O^{\text I}_\Delta(\bm x,t),\hat O^{\text I}_\Delta(\bm x',t')]\rangle_{\lambda_i}+\cdots,
 \end{split}
\end{gather}
where $\hat O^{\text I}_\Delta(\bm x,t)=\exp[i\hat H(t_0) (t-t_0)]\hat O_\Delta(\bm x)\exp[-i\hat H(t_0) (t-t_0)]$ and the expectation values on the RHS are evaluated in the initial ground state $\lambda_i$.
Due to the symmetries of the action, the two-point function behaves as (for the detailed calculation, see Supplemental material A)
\begin{gather}
 \begin{split}
 \label{e15}
&\langle [\hat O^{\text I}_\Delta(\bm x,t),\hat O^{\text I}_\Delta(\bm x',t')]\rangle_{\lambda_i}\\
&=\frac{1}{|\bm x-\bm x'|^{2\Delta}}g\left(\frac{|\bm x-\bm x'|^z}{t-t'},\lambda_i|\bm x-\bm x'|^{d-1-\Delta+z}\right).
 \end{split}
\end{gather}

Since the quench process is inside the impulse stage, the characteristic time should be $\delta$. Then according to Eqs.~(\ref{e1}, \ref{e3}), the characteristic length and energy read: $\xi_{\text Q}\sim \delta^{1/z}$, $E_{\text Q}\sim\delta^{-1}$. It follows from Eq.~(\ref{e15}) that
\begin{gather}
 \begin{split}
 \label{e16}
&g\left(\frac{|\bm x-\bm x'|^z}{t-t'},\lambda_i|\bm x-\bm x'|^{d-1-\Delta+z}\right)\\
&=g\left(\frac{|\bm x-\bm x'|^z}{t-t'},0\right)+O\left(\lambda_0\xi_{\text Q}^{d-1-\Delta+z}\right),
 \end{split}
\end{gather}
and
\be
 \label{e17}
\langle O_\Delta\rangle_{\text{re}}\sim\lambda_0\delta\xi_{\text Q}^{d-1-2\Delta}\sim\delta^{\frac{d-1-2\Delta+z}{z}}.
\ee
Thus, the results in Refs.~\cite{un2014,un2015} are special case ($z=1$) of ours.
Finally for work statistics,
following the similar procedure, we have (see Supplemental material B)
\be
\label{e18}
(\kappa_n)_{\text{re}}\sim \delta^{\frac{d-1-2\Delta+2z-nz}{z}}.
\ee
We would like to emphasize that this approximation is invalid when $\lambda_0\delta^{(d-1+z-\Delta)/z}>1$, which is consistent with the condition $E_0\delta>1$ since $E_0\sim|\lambda_0|^{z\nu}$ and $\nu(d-1+z-\Delta)=1$ (see Supplemental material A).

\begin{figure}[t]
\centering
\includegraphics[width=0.45\textwidth]{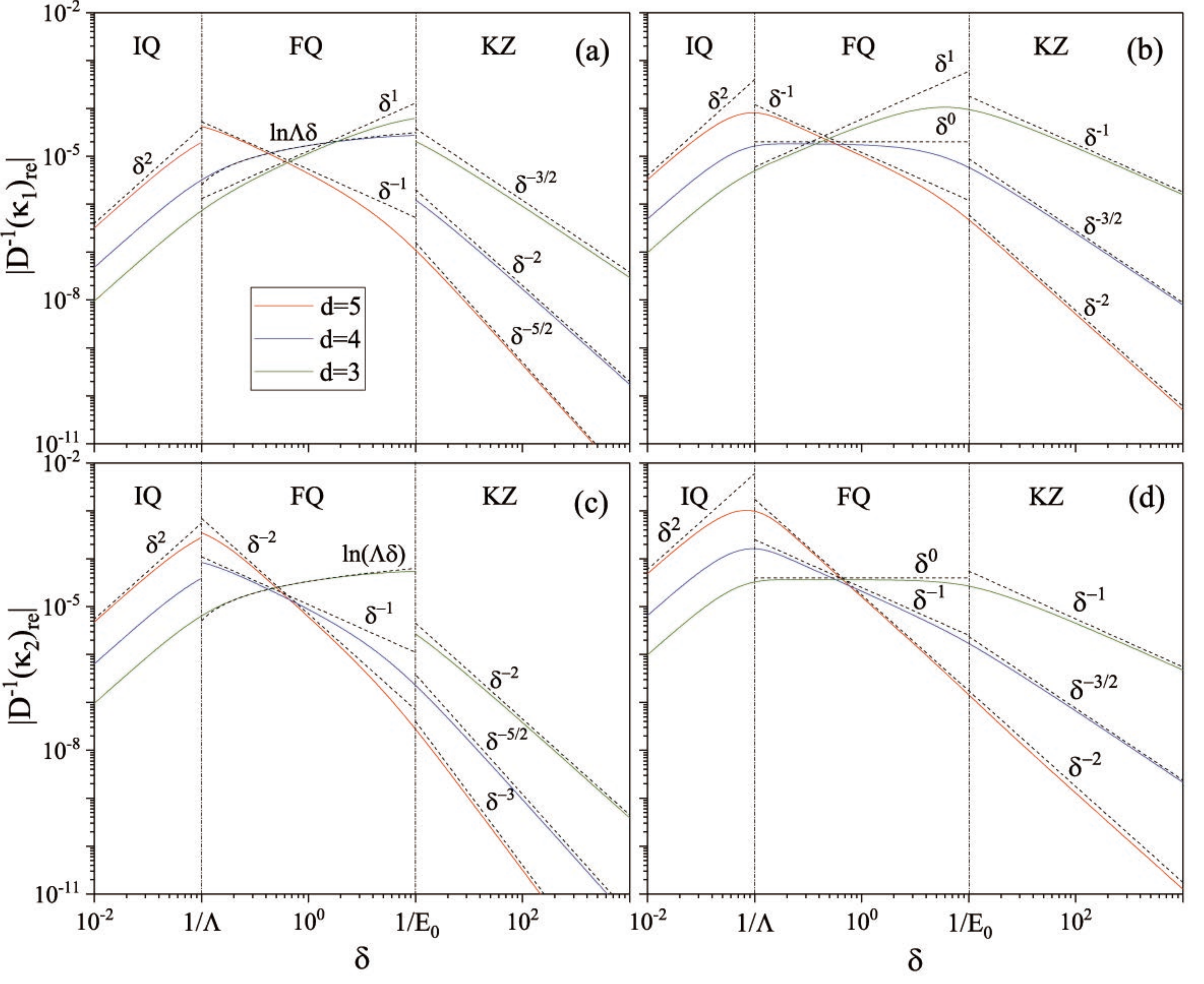}
\caption{The first and second cumulants of the work (absolute value) as a function of the quenching rate for a scalar field with time-dependent mass.  The parameters are: $E_0=m=0.1$, $\Lambda=10$. (a) and (c): $t_f=0$. (b) and (d): $t_f\to\infty$. The solid lines denote $(\kappa_n)_{\text{re}}$ in the Kibble-Zurek regime and fast-quench regime and $(\kappa_n)_{\text{su}}-\kappa_n$ in the instantaneous quench regime. The dashed lines denote the fitting curves. }\label{fig2}
\end{figure}

\emph{C: Instantaneous Quench}---The fast quench scaling is applicable when only the low-energy modes are excited, i.e., $\Lambda\gg E_0$. When $\Lambda\ll E_0$, the quench rate is fast compared to all physical scales. The evolution of the system can be approximated by its short-time solution and we call this the instantaneous quench regime (IQ). Then, under some conditions, we have~(Supplemental material C)
\be
\langle O_\Delta\rangle-\langle O_\Delta \rangle_{\text{su}}\sim\delta^2,
\ee
\be
\label{e20}
\kappa_n-(\kappa_n)_{\text{su}}\sim\delta^2,
\ee
where the subscript ``su'' denotes quantities in the sudden quench limit.

In short, our above analysis shows universal scaling behaviors of work statistics (Eqs.~(\ref{e11}, \ref{e12}, \ref{e18}, \ref{e20})) in three different regimes (KZ, FQ and IQ regimes). This is illustrated in Fig.~\ref{fig2} by the exact solution of the scalar field with changing mass, which is also studied analytically in the following. In the KZ regime, the frozen time determined the universal scaling behavior. And, work statistics exhibits different scaling behaviors in the two cases (quantum quench ends in the impulse stage or the adiabatic stage), while $\hat{O}_\Delta$ exhibits the same scaling behavior. The reason is that the energy gap is $\lambda$-dependent while $\hat{O}_\Delta$ is not. The change from the KZ regime to the FQ regime is due to the competition between the frozen time and the quench time (see Fig.~\ref{fig1}a). In the FQ regime, as a result of the symmetries of the anisotropic conformal field, the universal scaling behavior is determined by the scaling dimension of the physical quantities. The above scaling behaviors are applicable when only the low-energy modes are excited, i.e., $\Lambda\gg E_0$, which is broken in the IQ regime.

\emph{Renormalization}---Due to the divergence (UV or IR) of the field theory, the physical quantities should be renormalized~\cite{foot4}. In this section, we discuss the renormalization of $\langle O_\Delta\rangle$ as an example. For other quantities, the procedure of renormalization is applied straightforwardly. When $E_0\ll\Lambda$, in the KZ and FQ regimes, $\langle O_{\Delta}\rangle$ is divided into the sum of four parts:
\be
\label{e21}
\langle O_{\Delta}\rangle=\langle O_{\Delta}\rangle_{\text{ad}}+\langle  O_{\Delta}\rangle_{\text{hi}}+\langle  O_{\Delta}\rangle_{\text{re}}+\langle  O_{\Delta}\rangle_{\text{ig}},
\ee
where $\langle O_{\Delta}\rangle_{\text{ad}}$ denotes the zeroth-order adiabatic contribution, $\langle O_{\Delta}\rangle_{\text{hi}}$ denotes the higher-order adiabatic contribution which is UV divergent, $\langle \hat O_{\Delta}\rangle_{\text{re}}$ is independent of $\Lambda$ (called \emph{renormalized quantity}), and $\langle  O_{\Delta}\rangle_{\text{ig}}$ vanishes when $\Lambda\to\infty$.

Due to the adiabatic perturbation theory~\cite{hi1988,be2008,le2010}, when $\lambda(t)-\lambda(t_i)\sim (t-t_i)^r/\delta^r$, and $\lambda(t)-\lambda(t_f)\sim (t-t_f)^r/\delta^r$, the leading order of $\langle O_{\Delta}\rangle_{\text{hi}}$ is $\delta^{-2r}$ ($\delta^{-r}$) if $\hat O_{\Delta}$ commutes with $\hat H(t_f)$ (or not). For the consistence with the last section, $\langle  O_{\Delta}\rangle_{\text{hi}}$ only appears when its order is lower than the order of $\langle O_{\Delta}\rangle_{\text{re}}$, i.e., $\Delta r\nu/(1+z\nu r)>r$ ($\Delta r\nu/(1+z\nu r)>2r$) in the KZ regime and $(2\Delta-d+1-z)/z>r$ ($(2\Delta-d+1-z)/z>2r$) in the FQ regime~\cite{le2010,no2011,wo2020,foot2}. The independence of $\langle O_{\Delta}\rangle_{\text{re}}$ on $\Lambda/E_0$ reflects the anisotropic conformal symmetries under translation, spatial rotation, and dilation, which enables us to obtain the universal scaling of renormalized quantities.

The above observation is valid when $d-1-2\Delta+z<0$. When $d-1-2\Delta+z\geq 0$, in the fast quench regime, the systems exhibits IR divergence~\cite{un2015,sm2015,an2017}. Hence, in contrast to Eq.~(\ref{e21}), we choose another type of renormalization
\be
\langle  O_{\Delta}\rangle=\langle  O_{\Delta}\rangle_{\text{su}}+\langle O_{\Delta}\rangle_{\text{in}}+\langle  O_{\Delta}\rangle_{\text{re}}+\langle  O_{\Delta}\rangle_{\text{ig}},
\ee
where $\langle  O_{\Delta}\rangle_{\text{in}}\sim \delta^2$ corresponds to the scaling behavior in the instantaneous regime. Also, for the consistence with the results in the last section, $\langle  O_{\Delta}\rangle_{\text{in}}\sim \delta^2$ only appears when its order is lower than the order of $\langle O_{\Delta}\rangle_{\text{re}}$, i.e., $(d-1-2\Delta+z)/z>2$~\cite{foot2}.

It is concluded that the divergent part of quantities is renormalized by subtracting its value in the adiabatic limit (sudden quench limit) for the UV renormalization (IR renormalization).

\begin{table*}[htpb]
 \centering
 \caption{The first and second cumulants of work for a free field with changing mass in the KZ and FQ regimes}
 \label{t1}
\scalebox{0.9}{
\begin{tabular}{|c|c|c|c|c|c|c|}
\hline
                                                                                                                                               & \multicolumn{3}{c|}{Kibble-Zurek regime}                                                                                                                                                                                                          & \multicolumn{3}{c|}{Fast quench regime}                                                                                                                                                                                                                                            \\ \cline{2-7}
                                                                                                                                               & $d=5$                                                                            & $d=4$                                                                            & $d=3$                                                                            & $d=5$                                                                & $d=4$                                                                                                   & $d=3$                                                                                                  \\ \hline
$D^{-1}(\kappa_1)_{\text{re}},\ t_f=0$                                      & $0.014m^{5/2}/\delta^{5/2}$ & $0.017m^2/\delta^2$             & $0.030m^{3/2}/\delta^{3/2}$ & $0.052m^4/\delta$                    & $-m^4[0.063\ln(\Lambda\delta)+0.025]$ & $-0.13m^4\delta$                                                      \\ \hline
$D^{-1}(\kappa_2)_{\text{re}},\ t_f=0$                                     & $0.032m^3/\delta^3$             & $0.030m^{5/2}/\delta^{5/2}$ & $0.039m^2/\delta^2$             & $0.064m^4/\delta^2$ & $0.10m^4/\delta$                                                        & $-m^4[0.13\ln(\Lambda\delta)+0.050]$ \\ \hline
$D^{-1}(\kappa_1)_{\text{re}},\ t_f\to\infty$ & $0.051m^3/\delta^2$             & $0.080m^{5/2}/\delta^{3/2}$ & $0.16m^2/\delta$                                & $0.010m^4/\delta$                    & $0.18m^4$                                                                              & $0.52m^4\delta$                                                       \\ \hline
$D^{-1}(\kappa_2)_{\text{re}},\ t_f\to\infty$ & $0.13m^4/\delta^2$            & $0.22m^{7/2}/\delta^{3/2}$  & $0.48m^3/\delta$                               & $0.016m^4/\delta^2$ & $0.21m^4/\delta$                                                       & $0.37m^4$ \\
\hline
\end{tabular}}
\end{table*}

\emph{Example}---A lot of insights into this problem can in fact be obtained by looking at field theories with time-dependent parameters whose time evolution is exactly solvable, e.g., a free scalar field with changing mass in the momentum space and the Schr\"{o}dinger picture
\be
\hat H(t)=\frac{V_{d-1}}{2(2\pi)^{d-1}}\int_{k<\Lambda} \mathrm d^{d-1}k [\hat P_{\bm k}^2+\omega_{ k}(t)^2\hat Q_{\bm k}^2]
\ee
with the canonical quantization $[\hat Q_{\bm k},\hat P_{\bm k'}]=i\delta^{d-1}(\bm k-\bm k')$,
where the relativistic dispersion relation reads $\omega_{k}(t)=\sqrt{k^2+m(t)^2}$  (speed of light $c=1$), $k^2=k_1^2+k_2^2+\cdots+k_{d-1}^2$, and $m(t)$ denotes the changing mass. Hence, the energy-gap protocol is $E_g(t)=\omega_{k=0}=m(t)$ and the critical point is at $m(t)=0$.

The characteristic function $\chi(u)$ of this system with the initial ground state is analytically solved and results can be found in Refs.~\cite{sd2008,ps2013,zy2019} (for detailed calculation, see Supplemental material D). Here, for a specific protocol
\be
\label{e30}
m(t)^2=m^2[1-\cosh^{-2}(t/\delta)],
\ee
we obtain an exact solution for arbitrary quench rates ($\lambda(t)=m(t)^2/m^2, z=1,\nu=1/2, r=2, \Delta=d-2$).  And for simplicity, we discuss two cases: (1) $t_i\to -\infty, t_f\to\infty$; (2)  $t_i\to -\infty, t_f=0$.

We list the analytical results of the first and second work cumulants (renormalized) in KZ and FQ regimes for $d=5,4,3$ in TABLE~\ref{t1}. Here, $D\equiv V_{d-1}\Omega_{d-2}/(2\pi)^{d-1}$, $\Omega_{d-2}\equiv 2(2\pi)^{(d-1)/2}/\Gamma[(d-1)/2]$  is the solid angle in $d-1$ spatial dimensions, $\Gamma(s)$ is the Gamma function. Moreover, in the IQ regime, the scaling behavior $\kappa_n-(\kappa_n)_{\text{re}}\sim\delta^2$ is also obtained (see Supplemental material D). In Fig.~\ref{fig2}, we show the exact results (solid lines) and the fitting curves (dashed lines) in different cases. These results all verify our predictions in Eqs.~(\ref{e11}, \ref{e12}, \ref{e18}, \ref{e20}).

We thank Yu Chen for helpful discussions. The work was supported from the
National Basic Research Program of China (Grant No. 2016YFA0301201),  National Natural Science Foundation of China (Grant No. 12088101, No. 11534002), NSAF (Grant No. U1930403, No. U1930402).

 \appendix
 \renewcommand{\theequation}{S.\arabic{equation}}

 \setcounter{equation}{0}
\begin{widetext}
 \section{Supplemental Material: Universal scaling of work statistics in conformal-field-theory models}
\subsection*{A: anisotropic conformal field theory}
In our paper, the anisotropic conformal field theory means the field action is invariant under anisotropic scale transformation, translation, spatial rotation which form a closed Lie algebra~\cite{no2008}. For the anisotropic scale symmetry, the field action $S_{\text{CFT}}[\phi,O_{\Delta}]$ is invariant under the following transformation
\be
\label{s1}
\bm x\to c\bm x, t\to c^zt, \phi\to c^{\Delta_\phi}\phi, O_{\Delta}\to c^{\Delta}O_{\Delta},
\ee
where $z$ is the dynamical exponent, $\Delta_\phi,\Delta$ are the scaling dimension of the fields $\phi,O_{\Delta}$ respectively~\cite{no2007,no2008}. In addition, for Eq.~(14), the total action $S[\phi,O_{\Delta}]$ is invariant if we also transform $\lambda$ as
\be
\label{s2}
\lambda\to c^{\Delta-d+1-z}\lambda.
\ee

If the functional integration measure also has these symmetries, we have the following transformation of the correlation function~\cite{no2007,co2012}
\be
\label{s3}
\langle \hat O^{\text I}_\Delta(c\bm x_1,c^zt_1)\hat O^{\text I}_\Delta(c\bm x_2,c^zt_2)\cdots\hat O^{\text I}_\Delta(c\bm x_n,c^zt_n)\rangle_{c^{\Delta-d+1-z}\lambda}=c^{-n\Delta}\langle \hat O^{\text I}_\Delta(\bm x_1,t_1)\hat O^{\text I}_\Delta(\bm x_2,t_2)\cdots\hat O^{\text I}_\Delta(\bm x_n,t_n)\rangle_{\lambda}
\ee
Since the total action $S[\phi,O_{\Delta}]$ (for fixed $\lambda$) is also invariant under translation and spatial rotation, we have~\cite{no2008,no2007,co2012}
\be
\label{s4}
\langle [\hat O^{\text I}_\Delta(\bm x,t),\hat O^{\text I}_\Delta(\bm x',t')]\rangle_{\lambda}=\frac{1}{|\bm x-\bm x'|^{2\Delta}}g\left(\frac{|\bm x-\bm x'|^z}{t-t'},\lambda|\bm x-\bm x'|^{d-1-\Delta+z}\right).
\ee
Thus, both the higher order corrections in Eq.~(15)~\cite{un2014} and Eq.~(17) are up to the order $O\left(\lambda_0\delta^{(d-1-\Delta+z)/z}\right)$. Moreover, from Eq.~(\ref{s1}), the correlation length behaves as
\be
\label{s5}
\xi\to c\xi.
\ee
Comparing Eqs.~(\ref{s2}, \ref{s5}) with Eq.~(2), we have $c=c^{-\nu(\Delta-d+1-z)}$, i.e., $\nu(d-1+z-\Delta)=1$ which is also found in Refs.~\cite{no2011,sc2009,qu2007}.

\subsection*{B: work statistics in the fast quench regime}

In the fast quench regime, by using the perturbation theory, from Eq.~(14), we have~\cite{no2020}
\be
\label{s6}
\ln\chi(u)=iu(\lambda_f-\lambda_i)V_{d-1}\langle O_{\Delta}\rangle_{\text c}+ iu(\lambda_f^2-\lambda_i^2)V_{d-1}\int_{-\infty}^{\infty}\frac{\mathrm d\omega}{2\pi}\frac{G^>_{\text c}(\omega)}{\omega}+V_{d-1}\int_{-\infty}^{\infty}\frac{\mathrm d\omega}{2\pi}\frac{1-e^{iu\omega}}{\omega^2}A(\omega)G^>_{\text c}(\omega)+O(\lambda^3),
\ee
where the subscript ``c'' denotes that the quantities are evaluated in the ground state $\lambda=0$ (critical point) and only connected diagrams are included,
\be
\label{s7}
A(\omega)=\left|\int_{t_i}^{t_f}\mathrm dt \dot{\lambda}(t)e^{i\omega t}\right|^2,
\ee
and
\be
\label{s8}
G^>_{\text c}(\omega)=\int_{-\infty}^{\infty}\mathrm ds G_c^>(s)e^{i\omega s},\ G^>_{\text c}(s)=(-i)^2\int \mathrm d^{d-1}x \langle \hat O^{\text I}(\bm x,t)\hat O^{\text I}(0,0)\rangle_{\text c}.
\ee

Then, from Eq.~(\ref{s7}), it is straightforward to obtain that
\be
\label{s9}
A(\omega)=A_0(\omega)\equiv
\begin{cases}
(\lambda_f-\lambda_i)^2 \bm{1}_{\{0\}}(\omega)& \text{adiabatic limit}\\
(\lambda_f-\lambda_i)^2 & \text{sudden quench limit}
\end{cases},
\ee
where
\be
\bm{1}_{\{0\}}(\omega)
\begin{cases}
1&\omega=0\\
0&\omega\neq 0
\end{cases}
\ee
is the indicator function. According to Eqs.~(22, 23) and Eqs.~(\ref{s6}, \ref{s9}), we have
\be
(\ln\chi(u))_{\text{re}}\approx V_{d-1}\int_{-\infty}^{\infty}\frac{\mathrm d\omega}{2\pi}\frac{1-e^{iu\omega}}{\omega^2}[A(\omega)-A_0(\omega)]G^>_{\text c}(\omega).
\ee
Finally, because $G^>_{\text c}(\omega)\sim \delta\xi_{\text Q}^{d-1-2\Delta}$ (Eqs.~(16, 17) and Eq.~(\ref{s8})), $\omega\sim \delta^{-1}$, and $A(\omega)\sim O(1)$ (Eq.~(\ref{s7})), we have
\be
(\ln\chi(u))_{\text{re}}\sim \delta^{\frac{d-1-2\Delta+2z}{z}}
\ee
and
\be
(\kappa_n)_{\text{re}}\propto V_{d-1}E_{\text Q}^{n-1}\lambda_0^2\delta\xi_{\text Q}^{d-1-2\Delta}\sim \delta^{\frac{d-1-2\Delta+2z-nz}{z}}.
\ee

For a scalar field with changing mass (Eq.~(24)), from Eq.~(25) and Eqs.~(\ref{s7}, \ref{s8}), we have
\begin{gather}
 \begin{split}
 \label{s14}
A(\omega)=\left|\int_{-\infty}^{t_f}\mathrm dt \dot{\lambda}(t)e^{i\omega t}\right|^2=16\left|\sum_{l=1}^{\infty}\frac{l^2}{l+i\omega\delta/2}(-\eta)^l\right|^2,
 \end{split}
\end{gather}
where $\eta=e^{2t_f/\delta}$, and
\be
G^>_{\text c}(\omega)=\frac{-Dm^4}{8}\int_0^\Lambda \mathrm dk k^{d-4}2\pi\delta(\omega-2k).
\ee
The series converges when $t_f<0$,  but our following results (except Eq.~(\ref{s19})) are still valid when $t_f>0$ by using analytic continuation.
Then, we obtain
\be
\label{s16}
(\ln\chi(u))_{\text{re}}= \frac{Dm^4}{2}\int_{0}^{\Lambda}\mathrm dk k^{d-6}(e^{2iuk}-1)\times
\begin{cases}
\sum_{l, l'}\frac{l^2l'^2(ll'+k^2\delta^2)}{(l^2+k^2\delta^2)(l'^2+k^2\delta^2)}(-\eta)^{l+l'}& d\geq5\\
\sum_{l, l'}\left[\frac{l^2l'^2(ll'+k^2\delta^2)}{(l^2+k^2\delta^2)(l'^2+k^2\delta^2)}-ll'\right](-\eta)^{l+l'}& d<5
\end{cases}.
\ee
Let $\Lambda/m\to \infty$, from Eq.~(\ref{s16}), we have
\be
(\kappa_1)_{\text{re}}=Dm^4\times
\begin{cases}
\frac{\pi\eta^2(\eta^3+5\eta^2-5\eta+15)}{30(\eta+1)^5}\delta^{-1}& d=5\\
\frac{-\eta^2}{(\eta+1)^4}\ln(\Lambda\delta)+C_1& d=4\\
\frac{\pi\eta^2(\eta-3)}{6(\eta+1)^3}\delta& d=3
\end{cases},
\ee
and
\be
\label{s18}
(\kappa_2)_{\text{re}}=2Dm^4\times
\begin{cases}
\left[\frac{\eta^2(\eta-1)^2}{(\eta+1)^6}\ln(\Lambda\delta)+C_2\right]\delta^{-2}& d=5\\
\frac{\pi\eta^2(\eta^3+5\eta^2-5\eta+15)}{30(\eta+1)^5}\delta^{-1}& d=4\\
\frac{-\eta^2}{(\eta+1)^4}\ln(\Lambda\delta)+C_1& d=3
\end{cases},
\ee
where
\begin{gather}
 \begin{split}
 \label{s19}
C_1=&\sum_{l,l'}\frac{ll'(l\ln l'+l'\ln l)}{l+l'}(-\eta)^{l+l'}\\
C_2=&-\sum_{l,l'}\frac{l^2l'^2(l\ln l+l'\ln l')}{l+l'}(-\eta)^{l+l'}.
 \end{split}
\end{gather}
When $t_f=0$, numerical calculation of Eq.~(\ref{s19}) shows $C_1=-0.025$ and $C_2=0.032$.
When $t_f\to\infty$, from Eq.~(\ref{s18}), we have
\be
(\kappa_2)_{\text{re}}=2Dm^4\times
\begin{cases}
C_2\delta^{-2}& d=5\\
\frac{\pi}{30}\delta^{-1}& d=4\\
C_1& d=3
\end{cases}.
\ee
Here, because
\be
\label{s21}
A(\omega)=\frac{\pi^2\omega^4\delta^4}{\sinh^2(\pi \omega\delta/2)}
\ee
in this case (Eq.~(\ref{s7})), the calculation of $(\kappa_2)_{\text{re}}$ shows $C_1=3\zeta(3)/(2\pi^2)$ and $C_2=15\zeta(5)/(2\pi^4)$.

\subsection*{C: short-time evolution}

In the instantaneous quench regime, the evolution of the system can be approximated by its short-time solution. When $\delta\to 0, t_1=s_1\delta, t_0=0$, the time evolution operator $\hat U(t_1,t_0)=\mathcal {T}e^{-i\int_{0}^{t_1}\mathrm dt[\hat H_0+\lambda(t)\hat H_1]}$ is approximated as
\be
\label{s22}
\hat U(t_1,t_0)=1-i\delta\int_{0}^{s_1}\mathrm ds [\hat H_0+\lambda(s\delta)\hat H_1]-\delta^2\int_{0}^{s_1}\mathrm ds\int_{0}^{s}\mathrm ds' [\hat H_0+\lambda(s\delta)\hat H_1][\hat H_0+\lambda(s'\delta)\hat H_1]+O(\delta^3),
\ee
where $\mathcal T$ denotes the time-ordered operator.
For further discussion, let $|E_n(t)\rangle$ denote the instantaneous eigenstate of the time-dependent Hamiltonian, $[\hat H_0+\lambda(t)\hat H_1]|E_n(t)\rangle=E_n(t)|E_n(t)\rangle$, $|E_0(t)\rangle$ denote the ground state. Then, we have
\be
\label{s23}
\langle E_n(t_1)|\hat H_1|E_0(t_0)\rangle=\frac{E_n(t_1)-E_0(t_0)}{\lambda_f-\lambda_i}\langle E_n(t_1)|E_0(t_0)\rangle.
\ee
Thus, let $\alpha_n=\langle E_n(t_1)|\hat U(t_1,t_0)|E_0(t_0)\rangle$ and $\alpha^0_n=\langle E_n(t_1)|E_0(t_0)\rangle$ denote the transition probability amplitude for the instantaneous quench and sudden quench, it follows from Eqs.~(\ref{s22}, \ref{s23}) that
\begin{gather}
 \begin{split}
\alpha_n=&\alpha_n^0\left\{1-i\delta s_1E_0(t_0)-i\delta\frac{E_n(t_1)-E_0(t_0)}{\lambda_f-\lambda_i}\int_{0}^{s_1}\mathrm ds[\lambda(s\delta)-\lambda_i]\right\}\\
&-\delta^2\int_{0}^{s_1}\mathrm ds\int_{0}^{s}\mathrm ds' \langle E_n(t_1)|[\hat H_0+\lambda(s\delta)\hat H_1][\hat H_0+\lambda(s'\delta)\hat H_1]|E_0(t_0)\rangle+O(\delta^3)
 \end{split}
\end{gather}
Thus, it is easy to check that if an operator $\hat A$ satisfies $\langle E_m(t_1)|\hat A|E_n(t_1)\rangle(\alpha^{0}_m)^*\alpha_n^0=\langle E_n(t_1)|\hat A|E_m(t_1)\rangle(\alpha^{0}_n)^*\alpha_m^0$ ($*$ denotes the complex conjugate), we have
\be
\langle A \rangle-\langle A \rangle_{\text{su}}\sim \delta^2,
\ee
where $\langle A \rangle_{\text{su}}=\langle E_0(t_0)|\hat A|E_0(t_0)\rangle$.

\subsection*{D: the characteristic function of work for a free scalar field with changing mass}

Because the Hamiltonian of the field (Eq.~(24)) is a quadratic form of $\hat{P}_{\bm k}$ and $\hat{Q}_{\bm k}$, we obtain the cumulant characteristic function of work $\ln\chi(u)$ by using the representation of Lie group~\cite{zy2019} as
\be
\label{se24}
\ln \chi(u)=\frac{D}{2}\int_{0}^{\Lambda} \mathrm dk k^{d-2} \{iu[\omega_{k}(t_f)-\omega_{k}(t_i)]-\ln[1+n_{k}-n_{k}e^{2iu\omega_{k}(t_f)}]\},
\ee
where $D\equiv V_{d-1}\Omega_{d-2}/(2\pi)^{d-1}$, $\Omega_{d-2}\equiv 2(2\pi)^{(d-1)/2}/\Gamma[(d-1)/2]$  is the solid angle in $d-1$ spatial dimensions, $\Gamma(s)$ is the Gamma function,
\be
\label{se25}
n_{ k}=\frac{\omega_{k}(t_f)}{4\omega_{k}(t_i)}\left[y_{k}(t_f)^2+\bar{y}_{k}(t_f)^2+\frac{\dot{y}_{k}(t_f)^2+\dot{\bar y}_{k}(t_f)^2}{\omega_{k}(t_f)^2}\right]-\frac{1}{2},
\ee
the overhead dot denotes the time derivative, and $y_{ k}(t),\bar{y}_{ k}(t)$ are the general solutions of the following equation
\be
\label{se26}
\ddot{y}(t)+\omega_{k}(t)y(t)=0
\ee
with the initial condition $\{y_{k}(t_i),\dot{y}_{k}(t_i),\bar y_{ k}(t_i),\dot{\bar y}_{k}(t_i)\}=\{1,0,0,\omega_{k}(t_i)\}$.  Then, we have
\begin{gather}
 \begin{split}
 \label{se27}
\kappa_1=&\mu+D\int_{0}^{\Lambda} \mathrm dk k^{d-2}\omega_{ k}(t_f)n_{ k},\\
\kappa_2=&D\int_{0}^{\Lambda} \mathrm dk k^{d-2}2\omega_{ k}(t_f)^2n_{ k}(1+n_{ k}),
 \end{split}
\end{gather}
where
\be
\mu=\frac{D}{2}\int_{0}^{\Lambda} \mathrm dk k^{d-2} [\omega_{k}(t_f)-\omega_{k}(t_i)].
\ee
It follows from Eq.~(\ref{se27}) that $n_{ k}$ denotes the average number of the excited bosons in mode $\bm k$ after quench. and $\mu$ denotes the work done without any excitation (i.e., zeroth-order adiabatic contribution).

For the specific protocol (Eq.~(25) with $t_i\to-\infty$), we obtain the following results:
\begin{enumerate}[fullwidth,itemindent=0em,label=(\arabic*)]
\item Kibble-Zurek regime ($m\delta\gg 1$). In this regime, the characteristic momentum is $k_{\text c}\sim \sqrt{m/\delta}\ll m$~\cite{qu2016}. Hence, from the exact solution of Eq,~(\ref{se26}) (Supplemental material E), when $t_f=0$, we have
\begin{gather}
 \begin{split}
 \label{se31}
n_{ k}\approx\frac{\sqrt{m\delta}e^{-3\pi q_{ k}^2/4}(e^{\pi q_{k}^2}+1)}{4\pi q_{k}\sqrt{q_{ k}^2+m\delta}}\left[\left|\Gamma\left(\frac{3}{4}+\frac{iq_{ k}^2}{4}\right)\right|^2+\frac{q_{ k}^2}{4}\left|\Gamma\left(\frac{1}{4}+\frac{iq_{ k}^2}{4}\right)\right|^2\right],
 \end{split}
\end{gather}
where $q_{ k}=k\sqrt{\delta/m}$. From Eqs.~(\ref{se31}) and let $\Lambda \sqrt{\delta/m}\to \infty$ (only low-energy modes can be excited), we have
\be
(\kappa_1)_{\text{re}}=D\left(\frac{m}{\delta}\right)^{d/2}\times
\begin{cases}
0.014&d=5\\
0.017&d=4\\
0.030&d=3
\end{cases},
\ee
and
\be
(\kappa_2)_{\text{re}}=D\left(\frac{m}{\delta}\right)^{(d+1)/2}\times
\begin{cases}
0.032&d=5\\
0.030&d=4\\
0.039&d=3
\end{cases}.
\ee
Moreover, due to the power-law decay of Eq.~(\ref{se31}) ($n_{k}\to 1/(64q_{k}^8)$ when $q_{ k}\to \infty$), we cannot let $\Lambda \sqrt{\delta/m}\to \infty$ when $(d-1+n)/2\geq 4$, which results in that $(\kappa_n)_{\text{hi}}\sim \delta^{-4}\ln(\Lambda \delta)$ when $(d-1+n)/2=4$ and $(\kappa_n)_{\text{hi}}\sim \delta^{-4}$ when $(d-1+n)/2>4$.

Similarly, when $t_f\to \infty$, according to Supplemental material D and Eq.~(\ref{se24}), we have
\be
n_{ k}\approx e^{-\pi q_{ k}^2},
\ee
and
\begin{gather}
 \begin{split}
 \label{se37}
\ln\chi(u)=iu\mu+\frac{D\Gamma[(d-1)/2]}{4\pi^{(d-1)/2}}\left(\frac{m}{\delta}\right)^{(d-1)/2}\text{Li}_{(d+1)/2}(e^{2ium}-1),
 \end{split}
\end{gather}
where $\text{Li}_{n}(s)=\sum_{l=1}^{\infty}s^l/l^n$ is the polylogarithm function. Hence, we have $(\kappa_n)_{\text{re}}\sim (m/\delta)^{(d-1)/2}$ for any $n$. Moreover, due to the exponential decay of $n_{ k}$ when $k\to \infty$, we can  let $\Lambda \sqrt{\delta/m}\to \infty$ all the time, i.e., $(\kappa_n)_{\text{hi}}=0$, which is consistent with the fact that $\mathrm{d}^n\lambda(t)/\mathrm{d}t^n=0$ for any $n$ when $t\to \pm \infty$.

\item Fast quench regime ($m\delta\ll 1\ll\Lambda\delta$). Here, the characteristic momentum is $k_{\text c}\sim m \ll \sqrt{m/\delta}$~\cite{qu2016}. For convenience of calculation, we use perturbation theory to calculate $\ln\chi(u)$ (see Supplemental material B). Then, we have: when $t_f=0$,
\be
(\kappa_1)_{\text{re}}=Dm^4\times
\begin{cases}
\frac{\pi}{60}\delta^{-1}& d=5\\
-\frac{1}{16}\ln(\Lambda\delta)-0.025& d=4\\
-\frac{\pi}{24}\delta& d=3
\end{cases},
\ee
and
\be
(\kappa_2)_{\text{re}}=2Dm^4\times
\begin{cases}
0.032\delta^{-2}& d=5\\
\frac{\pi}{60}\delta^{-1} & d=4\\
-\frac{1}{16}\ln(\Lambda\delta)-0.025& d=3
\end{cases};
\ee
when $t_f\to\infty$,
\be
(\kappa_1)_{\text{re}}=Dm^4\times
\begin{cases}
\frac{\pi}{30}\delta^{-1}& d=5\\
\frac{3\zeta(3)}{2\pi^2}& d=4\\
\frac{\pi}{6}\delta& d=3
\end{cases},
\ee
and
\be
(\kappa_2)_{\text{re}}=2Dm^4\times
\begin{cases}
\frac{15\zeta(5)}{2\pi^4}\delta^{-2}& d=5\\
\frac{\pi}{30}\delta^{-1} & d=4\\
\frac{3\zeta(3)}{2\pi^2} & d=3
\end{cases},
\ee
where $\zeta(s)$ is the Riemann zeta function. Also, $(\kappa_n)_{\text{hi}}\sim \delta^{-4}\ln(\Lambda\delta)$ appears when $d-3-n=4$ and $(\kappa_n)_{\text{hi}}\sim \delta^{-4}$ appears when $d-3-n>4$. This is due to the fact that when $t_f=0$ and $\omega\to\infty$, $A(\omega)\to \omega^{-4}$ (Eq.~(\ref{s14})).

\item Instantaneous quench regime ($\Lambda\delta\ll 1$). In this regime, according to Supplemental material E, we have: when $t_f\to\infty$,
\be
\label{se41}
n_k\approx\frac{m^4\delta^2}{k^2+m^2};
\ee
when $t_f=0$,
\be
\label{se42}
n_k\approx\frac{(k-\sqrt{k^2+m^2})^2+(1-2\ln 2)m^4\delta^2}{4k\sqrt{k^2+m^2}}.
\ee
Hence, from Eqs.~(\ref{se41}, \ref{se42}), we have $\ln\chi(u)-(\ln\chi(u))_{\text{su}}\sim\delta^2$ and $\kappa_n-(\kappa_n)_{\text{su}}\sim\delta^2$.

\end{enumerate}

\subsection*{E: exact solutions of Eq.~(\ref{se26})}

According to Ref.~\cite{un2015,qu2016}, the general solutions of Eq.~(\ref{se26}) (with the protocol Eq.~(25)), $u_{ k}(t)$ and $u^*_{ k}(t)$, read (when $t_i\to-\infty$)
\be
u_{ k}(t)=\frac{2^{i\omega_{ k}(t_i)\delta}[\cosh(t/\delta)]^{2\alpha}}{B_1'B_2-B_1B_2'}\left[B_2\,{}_2F_1\left(a,b;\frac{1}{2};-\sinh^2(t/\delta)\right)+B_1\sinh(t/\delta)\,{}_{2}F_1\left(a+\frac{1}{2},b+\frac{1}{2};\frac{3}{2};-\sinh^2(t/\delta)\right)\right]
\ee
with the initial conditions: when $t\to -\infty$, $\{u_{k}(t), u^*_{ k}(t)\}\to \{e^{-i\omega_{ k}(t_i)t},e^{i\omega_{k}(t_i)t}\}$, where
\begin{gather}
 \begin{split}
B_1&=\frac{\Gamma(1/2)\Gamma(b-a)}{\Gamma(b)\Gamma(1/2-a)},\qquad \quad B_1'=\frac{\Gamma(1/2)\Gamma(a-b)}{\Gamma(a)\Gamma(1/2-b)},\\
B_2&=\frac{\Gamma(3/2)\Gamma(b-a)}{\Gamma(b+1/2)\Gamma(1-a)}, \quad \ B_2'=\frac{\Gamma(3/2)\Gamma(a-b)}{\Gamma(a+1/2)\Gamma(1-b)},\\
a&=\alpha+i\omega_{ k}(t_i)\delta/2, \qquad \quad \quad \;b=\alpha-i\omega_{ k}(t_i)\delta/2,\\
\alpha&=\frac{1-\sqrt{1-4m^2\delta}}{4},
 \end{split}
\end{gather}
and
\be
{}_2F_1(a,b,c;s)=\sum_{n=0}^{\infty}\frac{(a)_n(b)_n}{(c)_n}\frac{s^n}{n!}
\ee
denotes the usual hypergeometric function, and $(x)_n = x(x + 1) \cdots (x+n-1)$ ($(x)_0 = 1$). Thus when $t\to 0$, we have
\be
u_{k}(t)\to \frac{2^{i\omega_{ k}(t_i)\delta}}{B_1'B_2-B_1B_2'}\left(B_2+\frac{B_1}{\delta}\right).
\ee
And when $t\to \infty$, we have
\be
u_{ k}(t)\to \frac{2^{i\omega_{ k}(t_i)\delta+1}B_1B_2}{B_1'B_2-B_1B_2'}e^{-i\omega_{ k}(t_i)\delta}+\frac{B_1'B_2+B_1B_2'}{B_1'B_2-B_1B_2'}e^{-i\omega_{ k}(t_i)\delta}.
\ee
Moreover from Eq.~(\ref{se25}), we have
\be
n_{k}=\frac{\omega_{ k}(t_f)}{4\omega_{ k}(t_i)}\left[|u_{ k}(t_f)|^2+\frac{|\dot{u}_{ k}(t_f)|^2}{\omega_{ k}(t_f)^2}\right]-\frac{1}{2}.
\ee

\end{widetext}


\begin{thebibliography}{99}
\bibitem{foot1} When $E_0>\Lambda$, the fast quench regime is replaced by the non-universal regime (NU) (Fig.~{\ref{fig1}b}), where $E_{KZ}>\Lambda, E_0\delta>1$ and we do not find universal scaling behavior here.
\bibitem{foot4} In condensed matter physics, the cut-off energy $\Lambda$ is finite, and the divergence does not exist. However, it is still necessary to do renormalization in this case, because the existense of finite $\Lambda$ breaks the sysmmetries of the field and only the $\Lambda$-independent (renormalized) quantities exhibit universal scaling behavior.
\end{thebibliography}
\end{document}